\documentclass[twocolumn, referee]{svjour3}       
\smartqed  
\usepackage{graphicx}
\usepackage{amsmath}
\usepackage{amsfonts}
\usepackage{bm}
\usepackage{graphicx}
\usepackage{subfigure}
\usepackage{natbib}
\usepackage{url} 

\def\spacingset#1{\renewcommand{\baselinestretch}%
{#1}\small\normalsize} \spacingset{1}

\begin{document}

\title{Neural Mixture Distributional Regression
}


\author{David R{\"u}gamer \and
        Florian Pfisterer \and
        Bernd Bischl
}

\authorrunning{R{\"u}gamer et al.} 

\institute{David R{\"u}gamer, Florian Pfisterer, Bernd Bischl \at
Chair of Statistical Learning and Data Science\\
    Department of Statistics, LMU Munich\\
              Ludwigstr. 33, 80539 Munich \\
              \email{david.ruegamer@stat.uni-muenchen.de}           
}

\date{Received: date / Accepted: date}

\maketitle

\begin{abstract}
We present neural mixture distributional regression (NMDR), a holistic framework to estimate complex finite mixtures of distributional regressions defined by flexible additive predictors. Our framework is able to handle a large number of mixtures of potentially different distributions in high-dimensional settings, allows for efficient and scalable optimization and can be applied to recent concepts that combine structured regression models with deep neural networks. While many existing approaches for mixture models address challenges in optimization of such and provide results for convergence under specific model assumptions, our approach is assumption-free and instead makes use of optimizers well-established in deep learning. Through extensive numerical experiments and a high-dimensional deep learning application we provide evidence that the proposed approach is competitive to existing approaches and works well in more complex scenarios. 
\end{abstract}

\section{Introduction}

Mixture models are used to model the distribution of various sub-populations or sub-classes on the basis of data that is only observed for the pooled population. Mathematically, each sub-population is represented by a probability distribution and the pooled observations by a mixture of these distributions. Usually this is done without knowledge of the true class membership of each observed datum. Many applications of mixture models thus tackle the problem of finding memberships in an unsupervised fashion with clustering being one of the most prominent examples \citep{McLachlan.2019}. Mixture models have also been used in deep learning, e.g., for clustering \citep{Viroli.2019}, to build generative models for images using Gaussian mixtures \citep{Oord.2014} or as a hybrid approach for unsupervised outlier detection \citep{Zong.2018}. In contrast, a mixture regression model constitutes a supervised learning task and defines a mixture of (conditional) models for the outcome of interest, where the mixture components are still unknown and thus considered as latent variables. Various other distinctions exist, including the difference between finite and infinite mixture models \citep{Rasmussen.2000} or parametric and non-parametric situations. In the parametric case, an explicit parametric distribution assumption is used for each of the mixture components, while non-parametric approaches consider a mixture of non-parametric densities, each estimated, e.g., by a kernel density estimator \citep{McLachlan.2019}. This work focuses on finite mixtures of parametric (distributional) regression models.

\subsection{Mixture Models, Related Concepts and Optimization} 

As for classical statistical regression models, the goal of \textit{mixtures of regressions} or \textit{mixture regression models} is to describe the conditional distribution of an outcome, conditional on set of features \citep{Viele.2002}. Mixtures of regressions have been first introduced by \citet{Quandt.1958} under the term \textit{switching regimes} and have been extensively studied especially in the econometrics literature \citep{Benaglia.2009}. 

Models can be estimated based on various techniques, with the EM-algorithm based on the Maximum Likelihood principle being the most prominent one. Other approaches include Bayesian methods such as MCMC algorithms \citep{Wood.2002}, subspace clustering or tensor methods for mixed linear regression \citep{Zhong.2016}, stochastic Riemannian optimization \citep{Hosseini.2019} and gradient descent approaches \citep{Tsuji.2003}. The objective function induced by mixture models, in general, is non-convex and exhibits saddle points or local minima \citep{Murphy.2012}. Approaches targeting local optimization (EM, gradient descent) are known to be vulnerable to local optima \citep{Chaganty.2013}. For mixtures of Gaussian regression models, various theoretical convergence results exist, showing desired learning behaviour under certain regularity conditions \citep{Zhong.2016, Li.2018}. 

Mixture of experts \citep{Jacobs.1991} is a special case of a categorical mixture of normal distributions that can be fitted, e.g., with a softmax gating network, a double loop approach with inner iteratively-weighted least squares (IRLS) loop, a gating network in combination with an EM loop \citep{Xu.1995} or deterministic annealing \cite{Rao.1997}. 
The direct extension of categorical mixtures of normal distributions to other distributions was introduced by \citet{Bishop.1994} as so-called mixture density networks. The idea of mixture density networks is to use neural networks to learn mixtures of regression models in a distributional learning manner by allowing each parameter to be learned by a neural network.  

\subsection{Deep Learning and Additive Regression Models}

Recently, various publications highlight the advantages of using neural networks and deep learning platforms for additive regression models or, vice versa, constructing neural networks with an additive predictor inspired by statistical regression models \citep{Agarwal.2020, Chen.2020, Ruegamer.2020, Yang.2020}. While neural networks have an almost unlimited amount of applications and are one of the most flexible optimization frameworks for predictive and generative tasks, additive models allow for interpretable models, mathematically well-defined uncertainty quantification and can be extended easily by well-studied statistical modeling techniques such as smoothing splines or other structured additive effects. In order to make structured effects identifiable in the presence of a deep neural network, \citet{Ruegamer.2020} introduced a distributional regression framework that separates the deep learning predictors from the  structured components and thereby allows for identifiability of structured components in deep neural networks.

\subsection{Our Contribution}

While estimating mixture models on the basis of the EM algorithm works well for smaller problems, high-dimensional settings in which the number of parameters are similar to or exceed the number of observations, are often infeasible. As neural networks with optimizers from the field of deep learning (short: ODL) can be trained on batches of data with a potentially larger number of network weights than number of observations, fitting more complex mixtures within a neural network is appealing. Estimating additive mixture models via neural networks has many further advantageous. It provides the possibility to
 1) estimate a large mixture of regression models; 
2) flexibly define mixtures of different distributions;
3) allow for penalization to regularize the weights in each model term but also the amount of mixtures;
4) make use of optimizers established in the field of deep learning to tackle the notoriously difficult likelihood surface with its potentially many local maxima;
5) fit models in high-dimensional settings;
6)    extend models to include unstructured data sources and deep neural network structures.
Our framework provides these advantages while also extending mixtures of simple regression models to the combination of more complex additive models, e.g., fitting a mixture of semi-parametric distributional regression models. We call this framework neural mixture distributional regression (NMDR).

Training mixtures of structured regression models as NMDR turns out to be as reliable as or even more reliable than estimation via classical approaches in many different settings with the additional complementary advantages listed above. To our knowledge, we are also the first to provide a systematic comparison of EM-based and ODL-based estimation of mixture regression models.

\subsection{Comparison to Existing Approaches} 

NMDR learns every parameter in the mixture of distributions by means of a simple additive model, a deep neural network, or both. Our framework thereby unites neural density networks with (distributional) regression approaches to estimate statistical mixture models, incorporates several extensions of neural density networks by \citet{Bishop.1994}, such as penalized smooth effects, penalization for sparsity in mixture components, semi-structured predictors and considerations for various other extensions. Our approach also comprises various frameworks proposed in the statistical community such as \citet{Leisch.2004, Gruen.2007, Stasinopoulos.2007} to estimate mixtures of linear, generalized linear, generalized additive or distributional regression, but with the possibility for various extensions relevant in deep learning, the flexibility of a network architecture and a potentially more robust estimation routine. In contrast to existing approaches, NMDR defines the mixture probabilities as trainable model parameters and thereby also allows to estimate the mixture probabilities themselves on the basis of an additive model predictor.\\ 

\noindent In the following, we present our model definition in Section~\ref{sec:meth}, which also introduces the corresponding architecture for neural network-based optimization, identifiability considerations, penalized estimation approaches and further extensions. 
We then demonstrate the framework's properties using extensive numerical experiments in a third section and it's capability to handle high-dimensional unstructured data use cases in an applicaiton section. We conclude with a discussion in Section~\ref{sec:discuss}. In the supplementary material we additionally provide a large comparison of optimizers available in deep learning and make available all codes for our numerical experiments and the application on Github. 

\section{Methodology} \label{sec:meth}

Our goal is to model the conditional distribution of ${Y} | \bm{x}$ where ${Y}$ is the outcome of interest generated from a mixture of parametric distributions $\mathcal{F}_m, m \in \{ 1,\ldots,M \} =: \mathcal{M}$ that are influenced by a set of features $\bm{x} \in \mathbb{R}^p$.

\subsection{Model Definition}

Suppose we have $n$ independent univariate observations, $\bm{y} = (y_1, \ldots, y_n)^\top \in \mathbb{R}^n$, from the conditional distribution, conditional on $n$ observed feature vectors $\bm{X} := (\bm{x}^\top_1, \ldots, \bm{x}^\top_n)^\top \in \mathbb{R}^{n \times p}$. We define our model for the conditional distribution of $Y$ by the conditional density
\begin{equation} \label{eq:model}
    f_{Y|\bm{x}}(y_i | \bm{x}_i, \bm{\psi}) = \sum_{m=1}^M \pi_m (\bm{x}_i) f_m(y_i|\bm{\theta}_m(\bm{x}_i)),
\end{equation}
a mixture of density functions $f_m$ from distributions $\mathcal{F}_m$ each with its own $k_m$ distribution parameters $\bm{\theta}_m = (\vartheta_{m,1}, \ldots, \vartheta_{m,k_m})^\top$. $\pi_m \in [0,1]$ are mixture weights with $\sum_{m=1}^M \pi_m = 1$, usually assumed to follow a multinomial or multinoulli (categorical) distribution.  The vector $\bm{\psi} = (\bm{\pi}^\top, \bm{\theta}^\top)^\top \in \mathbb{R}^{M+K}$ with $K=\sum_{m=1}^M k_m$ comprises all parameters of the mixture distribution, namely $\bm{\pi} = (\pi_1,\ldots,\pi_M)^\top$ and $\bm{\theta} = (\bm{\theta}_1, \ldots, \bm{\theta}_M)^\top$. Each of the parameters $\psi_j, j=1,\ldots,(M+K)$ in turn is allowed to depend on features $\bm{x}_i$ through an additive predictor $\eta_j$ and a monotonic and differentiable function $h_j$, i.e.,
\begin{equation}
    \psi_j = h_j(\eta_j(\bm{x}_i)).
\end{equation}
The parameter-free transformation function $h_j$ ensures the correct domain of each $\psi_j$. For example, $h_j$ can be taken as the softmax function for the entries of $\bm{\psi}$ that correspond to the probabilities $\bm{\pi}$, ensuring $\psi_j \in [0,1]$ (see also \eqref{eq:pis}) . $\eta_j(\bm{x}_i)$ is the additive model part that ensures interpretability of the parameter $\psi_j$. For example, if $\psi_j$ corresponds to some $\pi_m$ and $\eta_j$ is a linear model, i.e., $\eta_j(\bm{x}_i) = \bm{x}_i^\top \bm{w}$, the network weights $\bm{w}$ can be interpreted as linear contributions of each of the feature to the logits of the mixture probability $m$. In general, all parameters in $\bm{\psi}$ are allowed to depend on some of the network's weights $\bm{w}$, but we omit the dependence on $\bm{w}$ for better readability. If the linear predictor comprises structured effects, the network's weights take the role of the coefficients in a corresponding regression models and can also be interpreted as such.

To train model (\ref{eq:model}), the empirical risk is used. This is given by the sum over all negative log-likelihood contributions of data points $i=1,\ldots,n$:
\begin{equation}
 \mathcal{R}_{emp}(\bm{\psi}) = -\sum_{i=1}^n \log \left\{ \sum_{m=1}^M \pi_m(\bm{x}_i) f_m(y_i|\bm{\theta}_m(\bm{x}_i)) \right\}, \label{eq:emprisk}
\end{equation}
which can be rewritten as sum of the log-likelihoods of both the model for the densities $f_m$ and the probabilities $\pi_m$ using the log-sum-exp function: 
\begin{equation}
    -\sum_{i=1}^n \log \left\{ \sum_{m=1}^M \exp \left[ \log \pi_m(\bm{x}_i) + \log f_m(y_i|\bm{\theta}_m(\bm{x}_i)) \right] \right\}. \label{eq:emprisk2}
\end{equation}
For $\pi_m$ we assume a softmax function $h$ to relate the linear predictor to the probabilities of the multinoulli distribution:
\begin{equation}
    \label{eq:pis}
\pi_m(\bm{x}_i) = \frac{\exp(\eta_m(\bm{x}_i))}{\sum_{l=1}^M \exp(\eta_l(\bm{x}_i))}, \quad \text{for } m \in \mathcal{M}.
\end{equation}
The loss of such mixture models can not be analytically optimized as the mixture probabilities depend on the parameters of the involved densities $f_m$ and the parameters of $f_m$, in turn, on the probabilities of the unobserved class memberships. The observed data log-likelihood is, in general, also neither convex nor concave and thus difficult to optimize. 
We 
therefore resort to optimizers from the field of deep learning by framing our model as a neural network, thereby also enabling various possible extensions presented in the following sections. Numerical experiments confirm that this choice -- when properly trained -- is not only more flexible and robust than EM-based optimization, but also yields feasible solutions for large data sets due to batch training. 

\subsection{Network Architecture}

\begin{figure*}
    \centering
    \includegraphics[page=6,  trim=1.5cm 2cm 2cm 0cm, width = 0.78\textwidth]{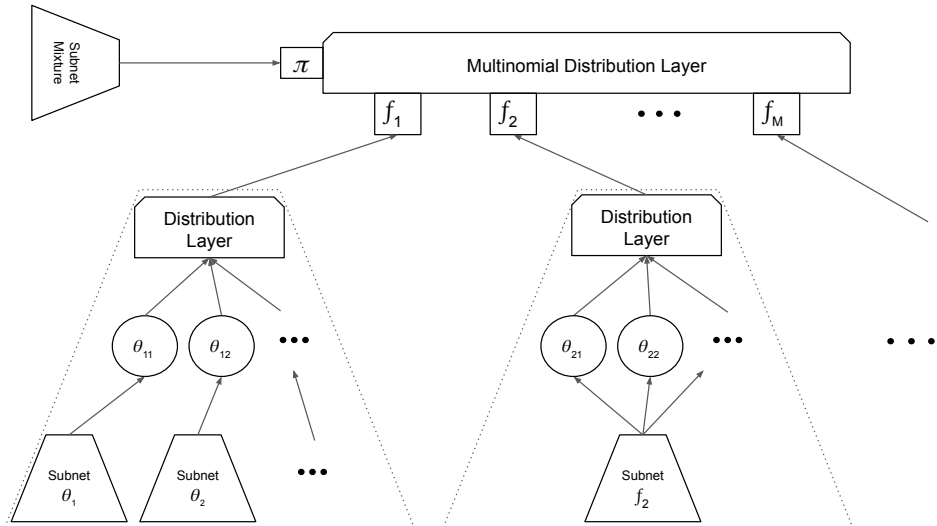}
    \caption{Example of an NMDR architecture. Smaller subnetworks (subnet) learn one or more parameters of a distribution which is defined in the respective distribution layer. For the first distribution in this example each distribution parameter in $\bm{\theta}_1$ is learned through a separate network while the second distribution is learned by a network that outputs all parameters $\bm{\theta}_2$ together. Each of the smaller distribution layers corresponds to a distributional regression model. The mixture model is then defined by an additional subnetwork that learns the mixture probabilities $\bm{\pi}$ as well as by the $M$ learned distributions $f_1, \ldots, f_K$.}
    \label{fig:arch}
\end{figure*}

We now describe the network architecture of NMDR. An exemplary architecture is depicted in Figure~\ref{fig:arch}. The network architecture implementing the above model (\ref{eq:model}) defines at most $K$ subnetworks, each modeling the linear predictor of a distribution parameter $\theta_{m,k}$. Using an appropriate output layer such as a softmax transformation function $h_j = \text{softmax}(\theta_j)$ for a parameter $\theta_j \in [0,1]$ or $h_j = \exp(\theta_j)$ for a non-negative parameter, these linear predictors are transformed into distribution parameters and passed to a distribution layer \citep{TFP.2017}. Each distribution layer corresponds to a mixture $f_m$ that is further passed to a multinomial or categorical distribution layer, modeling the mixture of all defined distributions. The mixture weights can either be directly estimated or also learned on the basis of input features using another subnetwork. A mixture of $M$ linear regressions, e.g., would be given by $M$ subnetworks, each learning the expectation of a normal distribution and a mixture subnetwork that only takes a constant input (a bias) and learns the $M$ mixture probabilities. 

NMDR is trained by optimizing empirical risk \eqref{eq:emprisk} as a function of the outcome $\bm{y}$ and the learned parameters $\bm{\psi}$. 
Using this architecture, both distribution components as well as mixture probabilities can either be learned by (a) structured models, such as linear or  additive models including smooth functions generated by a basis expansion, (b) a custom (deep) neural network or (c) a combination thereof. This includes special cases such as mixtures of linear regressions \citep{Viele.2002}, mixtures of semi-parametric regression models \citep{Stasinopoulos.2007} and mixtures of deep neural networks \citep{Tuske.2015}. 

\subsection{Identifiability}
Mixture models have several identifiability issues that we want to address in the following in order to define the scope to which NMDR can be successfully applied to. In general, a parametric family of distributions $\mathcal{F}(\bm{\psi}), \bm{\psi} \in \bm{\Psi}$ defined through density functions $f(y|\bm{\psi})$ as in \eqref{eq:model} is only identifiable \citep{Fruehwirth.2006} if for $\bm{\psi},\bm{\psi}^\dagger \in \bm{\Psi}$ it holds $$f(y|\bm{\psi}) = f(y|\bm{\psi}^\dagger) \text{ for almost all } y \in \mathcal{F} \Rightarrow \bm{\psi} = \bm{\psi}^\dagger.$$ Several more specific identifiability issues can be derived from this definition. 

\paragraph{Mixture Identifiability} If all $f_m$ belong to the same distribution family, e.g., $\mathcal{F}_m := \mathcal{N}(\vartheta_{m,1}, \vartheta_{m,2}) \, \forall m \in \mathcal{M}$, the parametrization (\ref{eq:model}) is only identifiable up to a permutation of the $M$ mixture components. This is also known as \textit{invariance to relabeling} \citep{Fruehwirth.2006}. A possible identifiability constraint can be imposed by enforcing an ordering of the mixture components $\pi_m$
$$0 \leq \pi_{1} \leq \pi_{2} \leq \ldots \leq \pi_{M}.$$ This constraint solves problems where probabilities comprise an order without ties, but still leads to an invariance of relabeling when there are at least two mixtures that are identical or when there is at least one mixture with $\pi_1 = 0$ \citep[which is also known as \textit{non-identifiability due to overfitting};][]{Fruehwirth.2006}. 
Note, however, that invariance to relabeling is mainly a problem when the parameters $\bm{\theta}_m$ of each density are either learned independently of features $\bm{x}$ or when $\eta_1 \equiv \ldots \equiv \eta_M$ and $\eta_{M+1} \equiv \ldots \equiv \eta_{M+K}$, i.e., no distinction is made between the additive predictors for each distribution. 
NMDR allows for the inclusion of monotonicity constraints through dedicated layers \citep[see, e.g.,][]{Gupta.2016} or self-defined transformation functions (e.g. by applying a softplus transformation in combination with a cumulative sum for the learned parameters). 

\paragraph{Generic Identifiability} 
Even if invariance of relabeling and non-identifiability due to overfitting can be ruled out, not all classes of distributions imply a generic identifiability \citep{Fruehwirth.2006}. That is, if
$$\sum_{m=1}^M \pi_m f(y_i|\bm{\theta}_m) = \sum_{l=1}^{L} \pi^\dagger_l f(y_i|\bm{\theta}^\dagger_l) \text{ for almost all } y \in \mathcal{F}$$ then $K=L$ and the two sets of parameters $\{\bm{\psi}\}$, $\{\bm{\psi}^\dagger\}$ are equivalent up to a reordering of their elements. Generic identifiability is, for example, violated, if two densities have the same functional form for different sets of parameters. A commonly known distribution to suffer from this problem is the binomial distribution $Bin(\varrho, \pi)$ when the total count of experiments $\varrho < 2M-1$ \citep{Teicher.1963}. For other distributions such as Gaussian, Poisson, Gamma, exponential or negative binomial, results of \citet{Yakowitz.1968} can be used to prove that these families of finite mixture distributions are generically identifiable.

Existing literature further requires the matrix $\bm{X}$ that is used in each additive predictor to have full rank \citep[see, e.g.,][]{Gruen.2007, Chaganty.2013} to obtain generic identifiability. Although this is theoretically relevant for NMDR, we note that high-dimensional settings with $p>n$ can be handled quite well as suggested by our experiments in Section~\ref{sec:numExp} and the subsequent deep learning application.

\paragraph{Additive Predictor Identifiability}

Identifiability issues can also occur on the level of the additive predictor. Especially for highly-correlated features, the interpretability of resulting models suffers from the non-identifiability of corresponding predictor weights. For example, $\eta_j = f_1(x_1) + f_2(x_2)$ can be equally obtained by using the functions $\tilde{f}_1$ and $\tilde{f}_2$, which are defined by adding and subtracting a constant $c$ to $f_1$ and $f_2$, respectively. We ensure identifiability by using sum-to-zero constraints for non-linear additive functions such as splines or tensor product splines and make use of the recently proposed orthogonalization cell \citep{Ruegamer.2020} if additive predictors are defined by both a structured additive predictor and a deep neural network for the same input covariate(s).

\subsection{Penalized Estimation of Structured Additive Predictors} \label{subsec:pen}

In order to estimate structured additive predictors such as splines within the neural network, the weights in each layer can be regularized using appropriate penalties or penalty matrices. When specified, these are simply added to \eqref{eq:emprisk}, i.e., 
$$
 \mathcal{R}_{pen}(\bm{\psi}) =  \mathcal{R}_{emp}(\bm{\psi}) + \rho \sum_{l \in \Lambda_1} |w_l| + \sum_{l \in \Lambda_2} \lambda_l \bm{w}_l^\top \bm{P}_l \bm{w}_l,
$$
with sets of weight indices $\Lambda_1$ determining the weights that are penalized using an $L_1$-penalty with complexity parameter $\rho$ and sets of index sets $\Lambda_2$ defining the weights that are penalized using a quadratic penalty with individual smoothing parameter $\lambda_l$ and individual penalty matrix $\bm{P}_l$. 

While estimating the tuning parameters is possible by including an inner optimization loop in the network, we use the approach suggested by \citet{Ruegamer.2020} to tune the different smooth effects 
by relating $\lambda_l$ to their respective degrees-of-freedom $\text{df}_l$. This has the advantage of training the network in a simple backpropagation procedure with only little to no tuning, by, e.g., setting the $\text{df}_l$ equal for all $l \in \Lambda_2$. 

\paragraph{Entropy-based Penalization of Mixtures} In order to allow for more flexibility in the estimation of mixture weights, neural networks and thus NMDR offers a plethora of possibilities. In Section~\ref{sec:highdimmix} we demonstrate the efficacy of NMDR when using a large mixture of distribution in the presence of only a few true underlying mixtures. To penalize an excessive amount of non-zero mixtures, we introduce an entropy-based penalty for the network weights that can be simply added to the objective function in NMDR:
\begin{equation}
    \mathcal{R}_{ent}(\bm{\psi}) = \mathcal{R}_{emp}(\bm{\psi}) - \xi \sum_m \pi_m \log \pi_m. \label{eq:entpen}
\end{equation}
The second part of \eqref{eq:entpen} is controlled via the tuning parameter $\xi \in \mathbb{R}$ and corresponds to the entropy induced by the categorical distribution of the (estimated) components in $\bm{\pi}$. A large value of $\xi$ forces the categorical distribution to be sparse in the amount of non-zero elements in $\bm{\pi}$, while smaller values will result in an (almost) uniform distribution of $\bm{\pi}$. As the entropy is permutation invariant w.r.t. the components in $\bm{\pi}$, this penalty is particularly suitable for mixtures that are only identified up to a permutation. We investigate the effects of this tuning parameter $\xi$ in section \ref{sec:highdimmix}.

\subsection{Further Extensions}

The NMDR framework can also be used to model various other problems. Classical mixture models for clustering mixtures in an unsupervised fashion is, e.g., a special case of the presented model when $\bm{\theta}_m(\bm{x}_i) \equiv \bm{\theta}_m$ and $\pi_m(\bm{x}_i) \equiv \pi_m \, \forall m \in \mathcal{M}$, i.e., when the parameters of the mixture distribution are not learned through additional features. NMDR can also be used for variable selection similar to \citet{Khalili.2007, Staedler.2010} by using network weight regularization for linear or additive features, or for distribution selection by using $M$ subnetworks to learn all mixture probabilities separately and regularize each subnetwork as done in the previous subsection to enforce a sparsity in the learned multinomial distribution. The extension to multivariate distribution is also straightforward and can be defined analogously, although training such mixtures is computationally more expensive. 

The presented approach can also be turned into a Bayesian neural network by using Bayesian layers \citep{TranGoogle.2018} which can be trained in the same manner using backpropagation with the \textit{Bayes by Backprop} \citep{Blundell.2015} approach. In this approach an additional loss term is added to the likelihood that incorporates prior distribution assumptions on layer weights. The resulting approximate posterior distribution of the network weights allows for statistical inference statements. 

\section{Numerical Experiments} \label{sec:numExp}

We now investigate our framework in terms of predictive performance and its capability to recover true underlying model coefficients and mixture probabilities. To this end, we first compare the NMDR with EM-based optimization routines to demonstrate the capability to keep up with state-of-the-art procedures. We then turn to misspecified mixture settings, in which we investigate situations where EM-based optimization fails and NMDR provides sparsity off-the-shelf. Further, we evaluate our approach in the presence of a mixture of generalized additive regression models with additional noise variables to demonstrate the framework's efficacy also outside of the distributional regression setting with more complex structured effects.

In general, we measure the estimation performance of model coefficients using the root mean squared error (RMSE), the goodness-of-fit of the estimated distribution using the log-scores (LS) and the prediction performance using the predicted log-scores \citep[PLS;][]{Gelfand.1994}, i.e., the empirical risk (\ref{eq:emprisk}) based on the estimated model parameters $\hat{\bm{\psi}}$ evaluated at the true outcome values of the training and test set for LS and PLS, respectively. We further evaluate the RMSE between the true and estimated mixture probabilities. For all measurements, we provide the mean and standard deviation over 20 simulation repetitions that are performed for each setting. If not stated differently, NMDR is ran with the \textit{rmsprop} \citep{rmsprop} optimizer with a learning rate of $0.1$ and cyclical learning rates \citep[CLR;][]{Smith.2017}. Optimization is run for 1500 epochs and a batch size of 50. An extensive analysis of several optimizers can be found in the supplementary material.

\subsection{Comparison with EM-based Optimization} \label{sec:comparisonEM}

We first compare the NMDR framework to an EM-based algorithm implemented in the R package \texttt{gamlss.mx} \citep{gamlss.mx} allowing for mixtures of various distributional regressions. We use $n \in \{300, 2500\}$ observations, $M \in \{2,3,5\}$ identically distributed mixture components, either following a Gaussian, a Laplace or a Logistic distribution, each defined by location and scale parameter and probabilities randomly drawn such that  $0.061 \leq \pi_m \leq 0.939$ (M=2) and $0.027 \leq \pi_m \leq 0.309$ (M=10), $p_m \in \{2, 10\}$ features for each distribution and distribution parameter in the mixture. While we test \texttt{gamlss.mx} with a fixed budget of $20$ restarts, we compare to NMDR using $1$ and $3$ restarts in order to assess the effect of multiple restarts. Each experimental configuration is replicated $10$ times.\\
\textbf{Results}: In the $p_m = 2$ scenario, NMDR consistently yields better log-scores, while mixing probabilities $\pi_m$ and coefficients $\theta$ are estimated more accurately by the EM based approach in the $n=300$ setting.
NMDR on the other hands improves over EM for $n=2500$.
The EM-based approach does not yield valid solutions for the $p_m = 10$ scenario, whereas NMDR does not suffer from high-dimensional settings and coefficient estimation is not deteriorated. Additional restarts for NMDR consistently improve estimation of coefficients, even though NMDR with only a single restart does only rarely suffer from local optima. 
Overall some outliers can be observed across all methods, suggesting a higher number of restarts might lead to further improved results.
\begin{figure}
    \centering
    \includegraphics[width = \columnwidth]{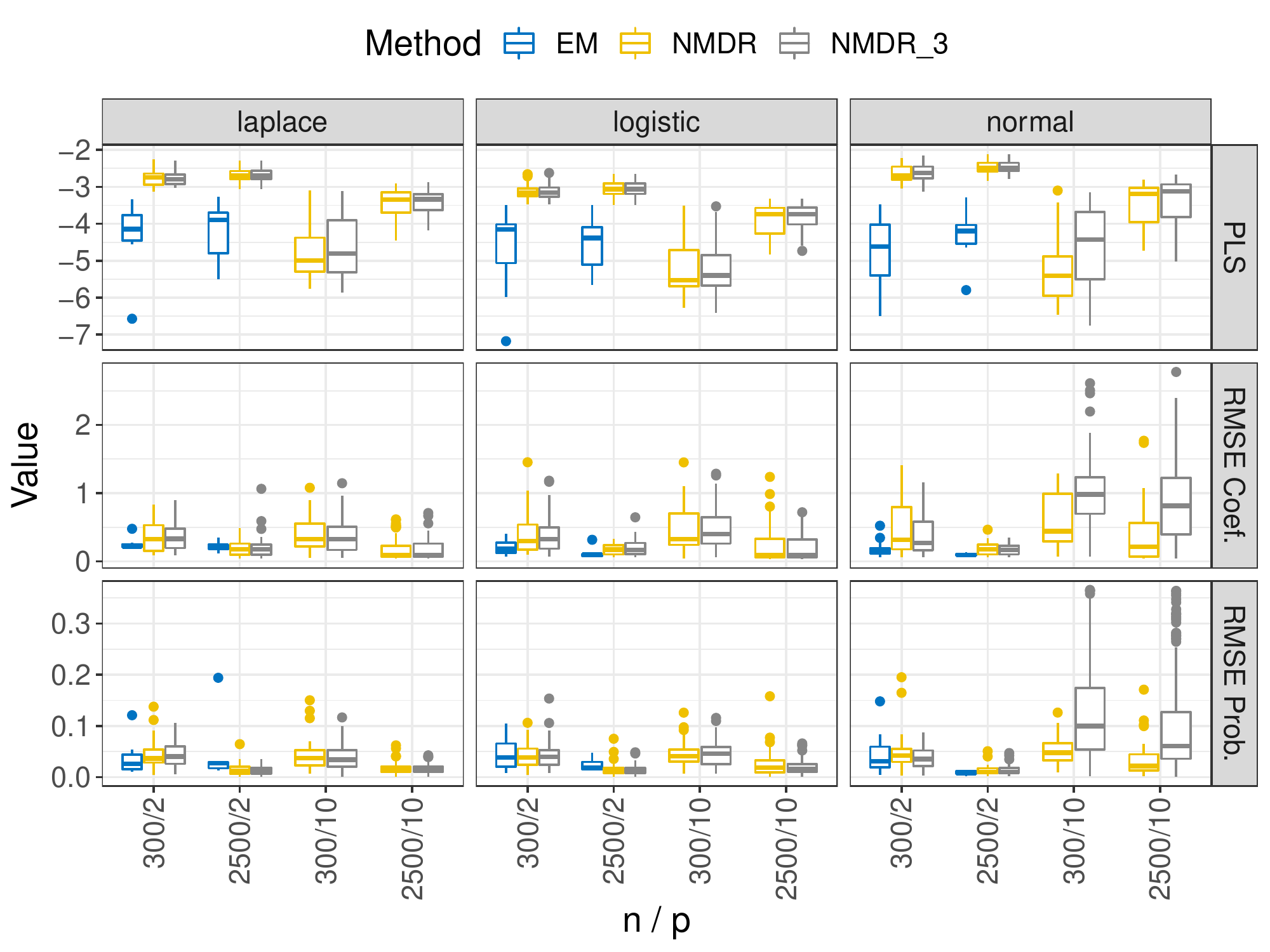}
    \caption{Comparison of EM-based optimization (EM), NMDR with one restart (NMDR) or with three restarts (NMDR\_3) for different distributions (columns), measures (rows) and combinations of $n$ and $p$. Missing boxplots of EM are due to missing valid solutions. RMSEs for coefficients $ > 3$ are omitted to improve readability.}
    \label{fig:comparison_em}
\end{figure}

\subsection{Misspecified Mixtures and Sparsity} \label{sec:highdimmix}

\begin{figure}
    \centering
    \includegraphics[width =\columnwidth]{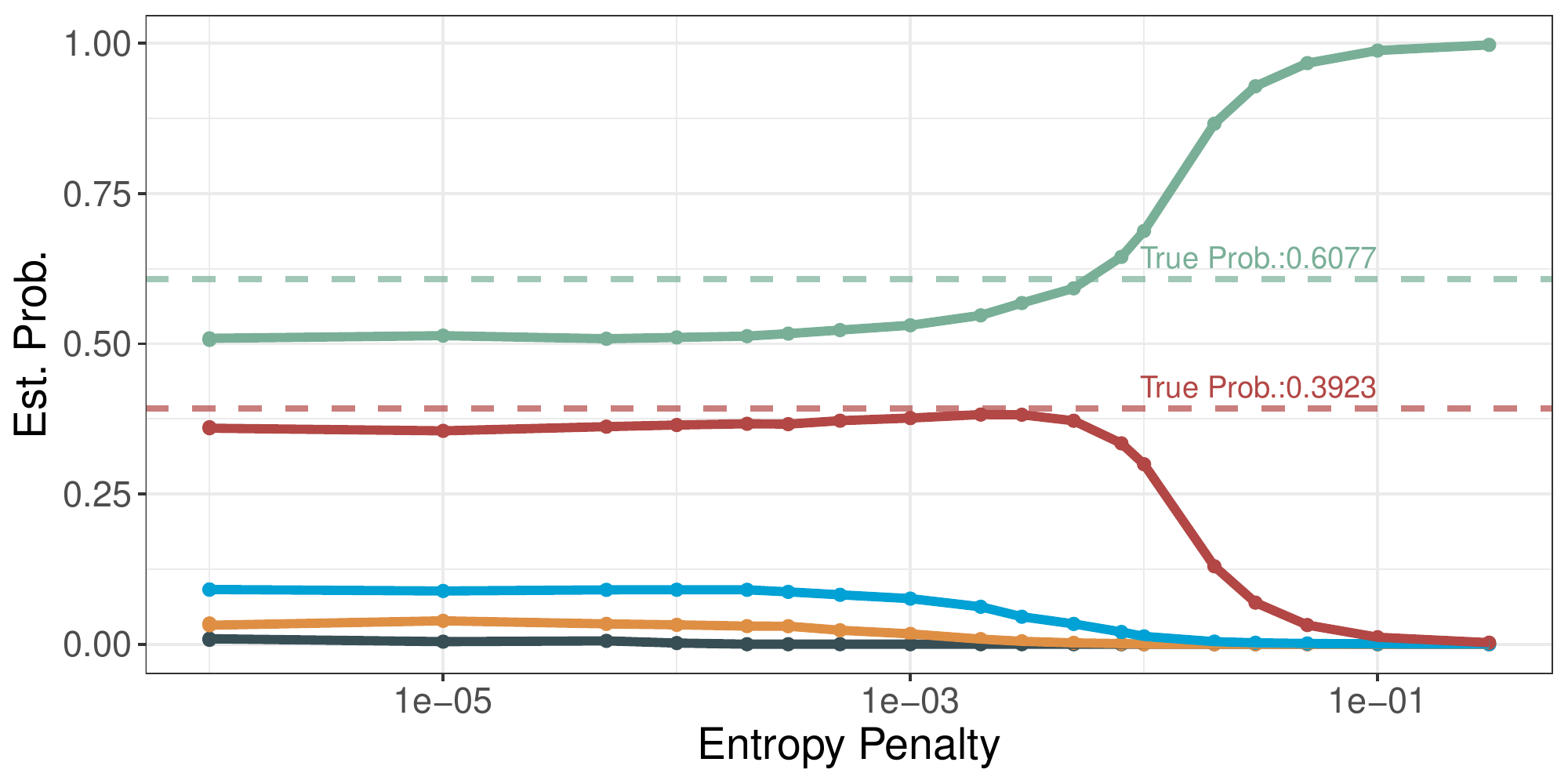}
    \caption{Coefficient path (estimated mixture probabilities) for different entropy penalties.}
    \label{fig:coef_path}
\end{figure}

In this simulation we use a normal mixture with $p_m = 10$ fixed predictors for each distribution and evaluate the estimation of mixture probabilities by NMDR for $n\in \{300, 2500\}$ when increasing the number of specified distributions $M^\dagger \in \{3, 5, 10\}$ by the assumed model for only $M=2$ actual mixture components. To allow for spareness in $\bm{\pi}$, we use the objective function $\mathcal{R}_{ent}$ introduced in Section~\ref{subsec:pen}.\\
\textbf{Results}: Results for various settings of the entropy penalty parameter $\xi$ are depicted in Figure~\ref{fig:add_mix} in the supplementary material. Correctly setting $\xi$ improves estimated parameters, yet the correctly specified model yields slightly more accurate results. We additionally investigate the coefficient path obtained from varying values of the penalty parameter $\xi$ for one simulated example.
Results for the $n=2500, M^\dagger = 2+3$ scenario are depicted in Figure \ref{fig:coef_path}. $\xi$ is varied between 0 and 1 on a logarithmic scale. The true model has two non-zero probabilities $0.6077$ and $0.39223$ while the other $3$ entries in $\bm{\pi}$ are 0.

\subsection{Mixture of Additive Regression Models} \label{sec:mixofdist}

\begin{figure}
    \centering
    \includegraphics[width = \columnwidth]{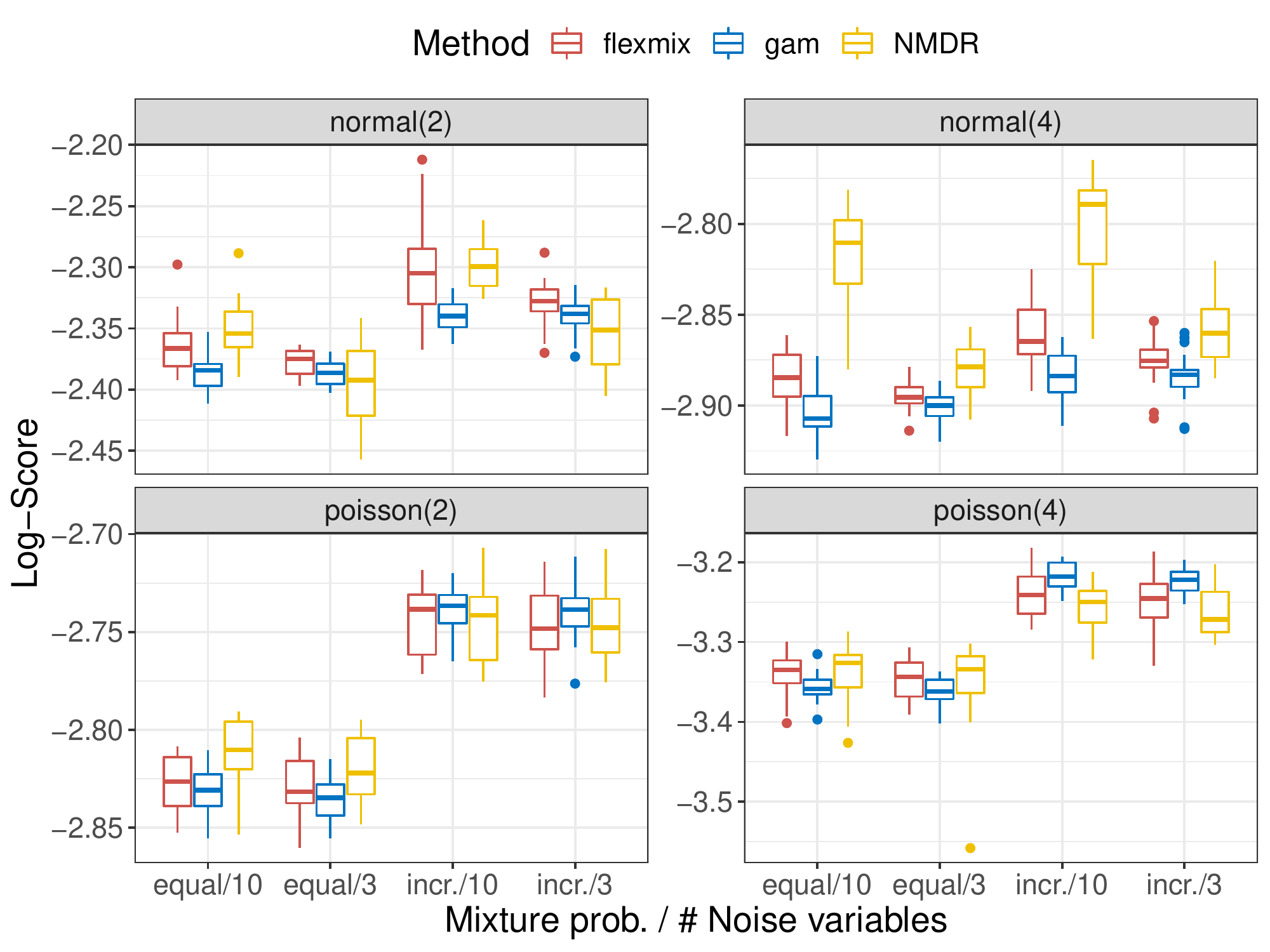}
    \caption{Comparison of average log-scores of a state-of-the-art implementation (flexmix), an oracle varying coefficient model (gam) and our approach (NMDR) in different colours for the two distributions (rows) and two scales (columns) as well as different mixture probabilities and number of noise variables (x-axis).}
    \label{fig:additive}
\end{figure}

Next we investigate mixtures of distributional regression models with non-linear effects in the linear predictors of the mean distribution parameter. We therefore generate $M = 3$ mixtures of a Poisson or normal distribution with expectation defined by $h(\beta_0 + f_1(x_1) + f_2(x_2) + f_3(x_3)$ with $f_1(x) = 2 \sin(3x)$, $f_2(x) = \exp(2x)$ and $f_3(x) = 0.2 x^{11} (10 (1 - x))^6 + 10 (10 x)^3  (1 - x)^{10})$. All covariates are independently drawn from a uniform distribution $\mathcal{U}(0,1)$. We model those effects using thin-plate regression splines from \citet{Wood.2017}. For Poisson data we use $h(\cdot) = \exp(\cdot)$ and the identity for the Gaussian case. $\bm{\pi}$ is either $(1/3,1/3,1/3)$ (``equal'' case) or $(1/10, 3/10, 6/10)$ (``inceasing'' case). We compare different number of noise variables $p_{Noise} \in \{3,10\}$ that are additionally modeled as non-linear predictors in the expectation parameter and investigate two different scale values $2$ or $4$, which either  define the Gaussian variance in each mixture component or a multiplicative effect in the linear predictor in the Poisson case. We compare our approach with an state-of-the-art implementation of additive model mixtures using the R package \texttt{flexmix} \citep{Leisch.2004} and, as a stable oracle reference, a generalized additive model with varying coefficients for all smooth effects, where the class label (unknown to the other two approaches) is used as the varying parameter. For NMDR, the smoothing parameters are determined as described in Section~\ref{subsec:pen} via the respective degrees-of-freedom which are set equally for all smooths to $10$ for normal and $6$ for Poisson distribution.\\ 
\textbf{Results}: The comparison for all settings is depicted in Figure~\ref{fig:additive} showing the log-scores divided by the number of observations $n$ for each of the approaches. Results suggest that our approach is competitive to both state-of-the-art and the oracle with no notable differences in many cases. NMDR shows better overall results for various cases but slightly higher variance of the results in some settings.




\section{Deep Mixture Distributional Regression} \label{sec:deepmix}

\begin{figure}
    \centering
    \includegraphics[width = \columnwidth]{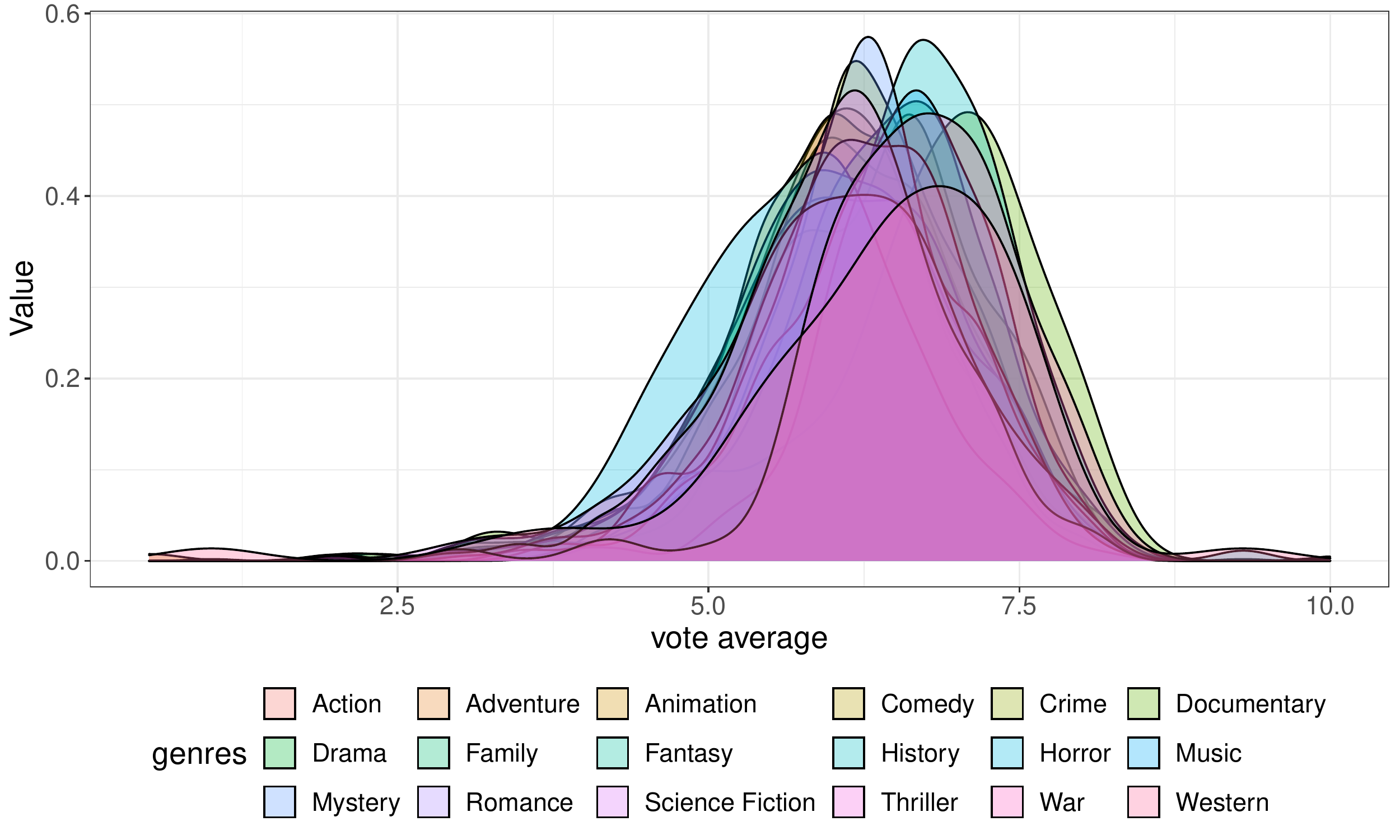}
    \caption{Distribution of movie genres in the movie data set.}
    \label{fig:genres}
\end{figure}

We now demonstrate the flexibility of our approach by extending and comparing a mixture of additive regression models with smooth terms by a deep neural network component that additionally feeds information about unstructured data into several additive predictors of NMDR. We use an ensemble of movie reviews from Kaggle consisting of $n=4442$ observations of (averaged) movie ratings that are available together with further covariates such as the movie's production budget, its popularity, the run time, the release date and a short description of the movie. The movies are an ensemble of genres, depicted in Figure~\ref{fig:genres}. 
where each movie can have one to several different genres. We therefore use the genre as latent variable to describe the mixture of distributions. After scaling the movie ratings to a scale from 0 to 1, we define a large mixture of 18 beta distributions in our analysis, where we model both distribution parameters $c_0, c_1$ of all 18 mixtures using the linear predictor
\begin{equation*}
\begin{split}
&s_{1,m,k}(budget_i) + s_{2,m,k}(popularity_i) +\\ 
\quad &s_{3,m,k}(runtime_i) + s_{4,m,k}(releasedate_i)        
\end{split}
\end{equation*}
with four different smooth terms $s_j(\cdot), j=1,\ldots,4$ for $m=1,\ldots,18$ mixture components and $k \in \{1,2\}$ distribution parameters. The mixture probabilities are estimated as constants. This baseline model (referred to as model (I)) is not further enhanced with text representations. In order to analyze the effect of the available textual description, we additionally investigate five different network definitions that include a deep neural network as input in the linear predictors of distribution parameters in addition to the structured predictors defined for the baseline model. The resulting networks all include an embedding layer of dimension $300$ with maximal sequence length of $100$ and maximum amount of $50000$ unique words. This results in additional 1.5 million parameters that have to be learned. We then flatten the embedding vectors and add 
\begin{itemize}
    \item[(II)] a dense layer with 18 units, where exactly one output of the dense layer is fed into one linear predictor of the shape parameter $c_0$ for each mixture;
    \item[(III)] a dense layer with 18 units, where exactly one output of the dense layer is fed into one linear predictor of the shape parameter $c_1$ for each mixture;
    \item[(IV)] a dense layer with 36 units, where exactly one output of the dense layer is fed into one linear predictor of the shape parameter $c_0$ or $c_1$ for each mixture;
    \item[(V)] a dense layer with 1 unit, which is fed into the linear predictor of $\bm{\pi}$;
    \item[(VI)] a dense layer with 37 units to combine modle (IV) and (V).
\end{itemize}
For model (VI) this, e.g., results in around 1.1 million additional trainable parameters. We set aside 20\% test data for final evaluation and train the models on 80\% of the data using a batch size of 256 and \textit{adadelta} as optimizer.\\
\begin{figure*}[h]
    \centering
    \subfigure[Comparison of different distributions of mixture components for each model. Each colour represents one mixture component.]{\includegraphics[width=0.48\textwidth]{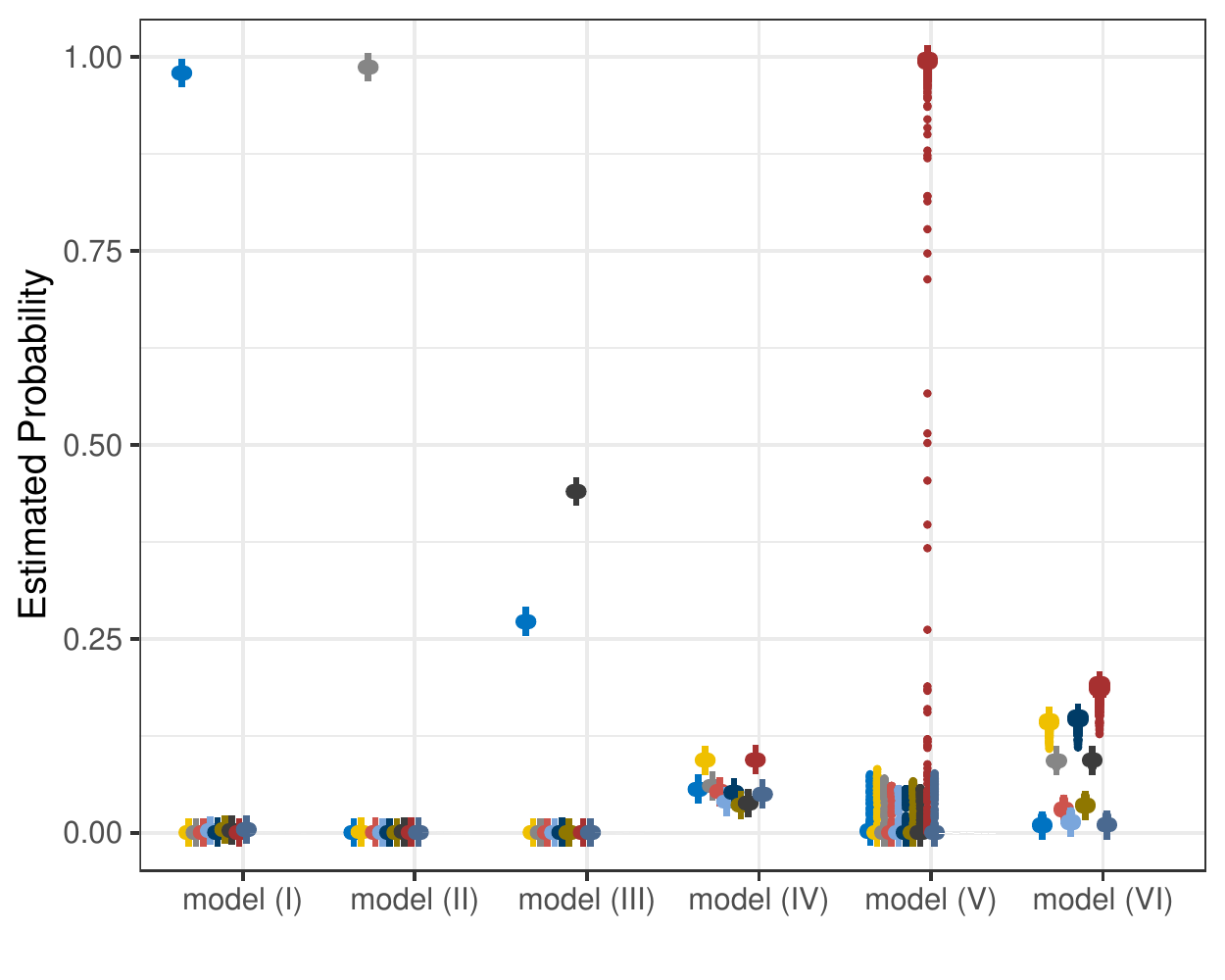}} 
    \quad
    \subfigure[t-SNE plot of the learned embedding space for model (V) for the 50 most common words.]{\includegraphics[width=0.48\textwidth]{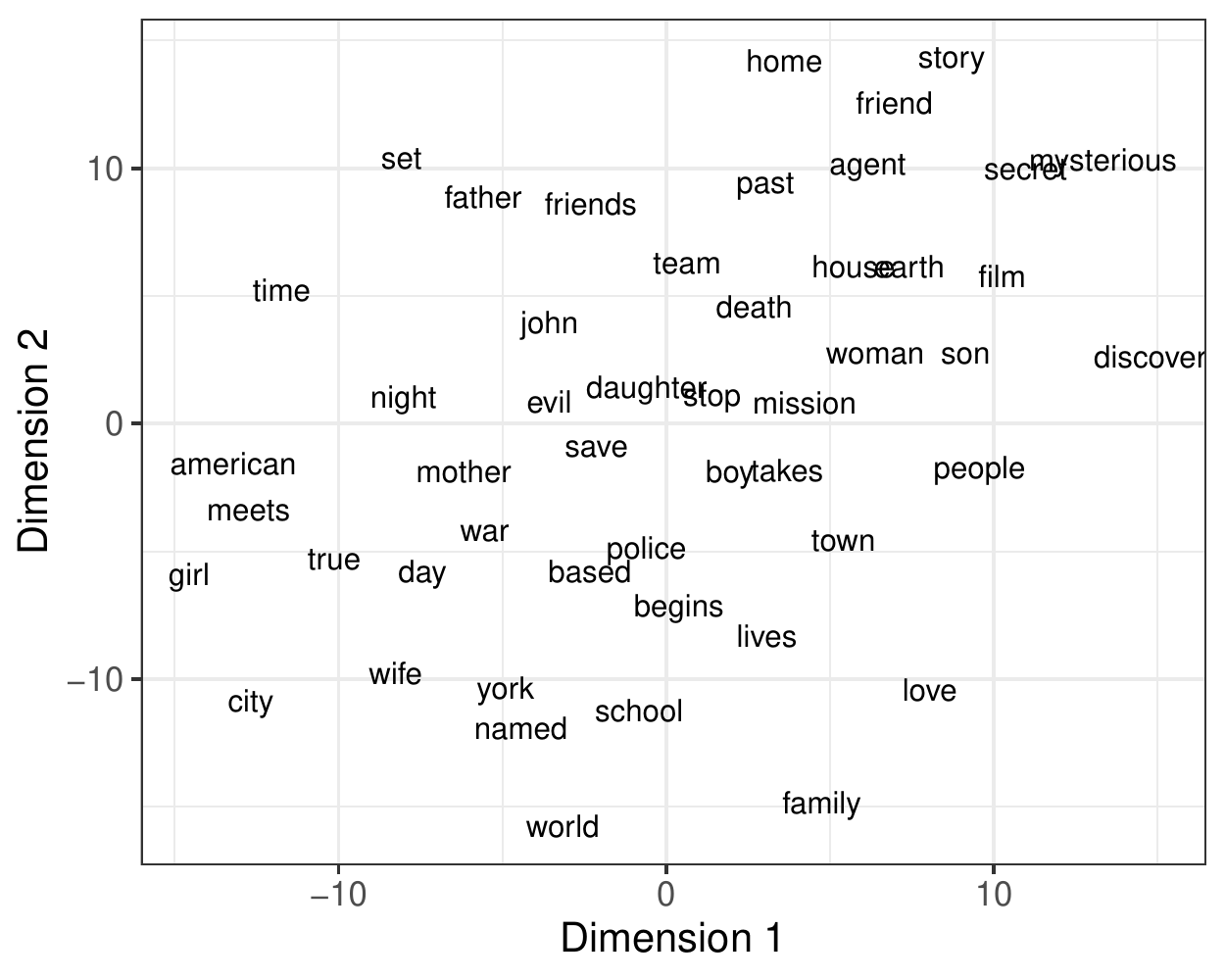}}
    \label{fig:modelcomp}
\end{figure*}
\textbf{Results}: All models show similar non-linear effects for all smooth terms while models with an added deep network tend to favor constant or linear effects over smooth effects for structured additive components. In contrast, estimated probabilities vary greatly across models with some models yielding a unimodal distribution of $\bm{\pi}$ while others have several non-zero mixture probabilities. We compare 
the estimated probabilities of all models in Figure~\ref{fig:modelcomp}. Table~\ref{tab:compModels} shows the mean root mean squared error (RMSE) for all models on the test data set when repeating the train-test split $10$ times. Here, model (V) with an observation-specific mixture performs notably better than all other models.
\begin{table}
\begin{small}
\begin{center}
\begin{tabular}{cc}
Model & Mean RMSE (std.)\\ \hline
(I) & 0.242 (0.128)\\
(II) & 0.176 (0.122)\\
 (III) & 0.213 (0.117)\\ 
 (IV) &  0.321 (0.156)\\
 (V) & \textbf{0.117} (0.026)\\
 (VI) & 0.190 (0.090) \\\hline
\end{tabular}
\caption{Mean RMSE values (standard deviation in brackets) of the different models on the test data set over 10 different train-test splits.}
\label{tab:compModels}
\end{center}
\end{small}
\end{table}
In addition, Figure~\ref{fig:modelcomp} visualizes the learned embedding space using a t-SNE plot in two dimensions.

\section{Discussion} \label{sec:discuss}

We have introduced the neural mixture distributional regression framework which combines statistical regression models with neural networks. In combination with state-of-the-art optimizers, this yields a competitive approach for learning mixture regression models. The flexibility and efficacy of neural network-based estimation of such models is competitive to existing approaches, which is demonstrated through numerical experiments. Results do not only highlight the ability of NMDR to achieve state-of-the-art performance, but also enable combinations with deep neural networks to allow for far more flexible approaches that go beyond a simple mixture of linear regressions.

\begin{acknowledgements}
This work has been partly funded by the German Federal Ministry of Education and Research (BMBF) under Grant No. 01IS18036A. The authors of this work take full responsibilities for its content.
\end{acknowledgements}

%
%

\bibliographystyle{spbasic}      

\spacingset{1}
\begin{small}

\bibliography{mybibfile}

\end{small}

\clearpage


\begin{appendix}

\section{Supplementary Material}

\subsection{Misspecified Mixtures and Sparsity} 

\begin{figure}[b]
    \centering
    \includegraphics[width = \columnwidth]{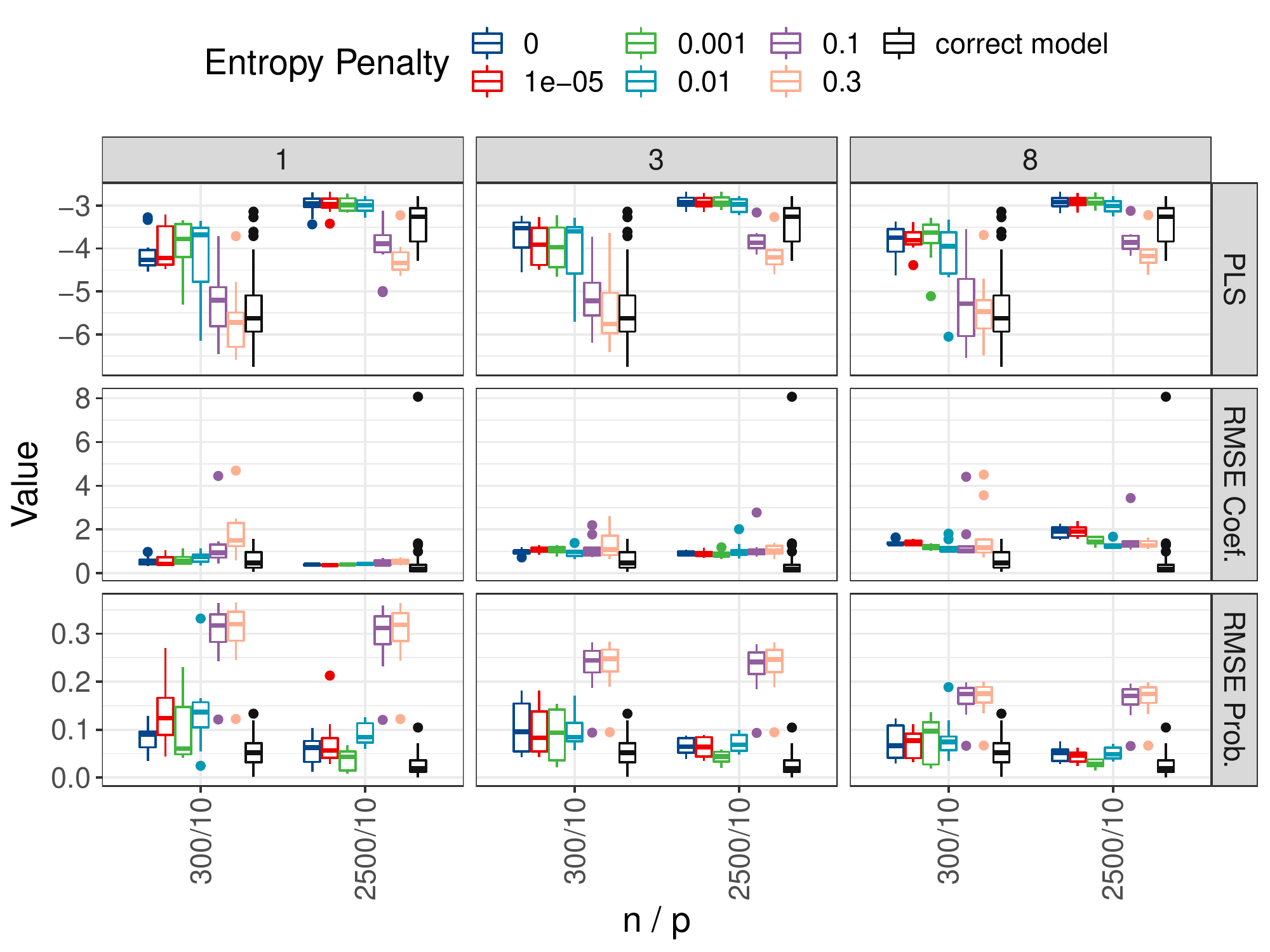}
    \caption{Model quality for misspecified models with specified mixtures $M^\dagger \in \{3, 5, 10\}$ (columns) instead of actual $M=2$ mixtures and different goodness of fit measures (rows). Colours correspond to different settings of $\xi$ or represent estimation results of the correct model (black).}
    \label{fig:add_mix}
\end{figure}

The folllowing Figure~\ref{fig:add_mix} shows the performance of NMDR defined with too many mixture components, which are then regularized through an entropy penalty. Results are depicted for different regularization values $\xi$ and compared them with an NMDR model with a correctly specified number of mixtures.

\subsection{Sensibility to hyperparameters} \label{app:comparison}

\begin{figure}[t]
    \centering
    \includegraphics[width = \columnwidth]{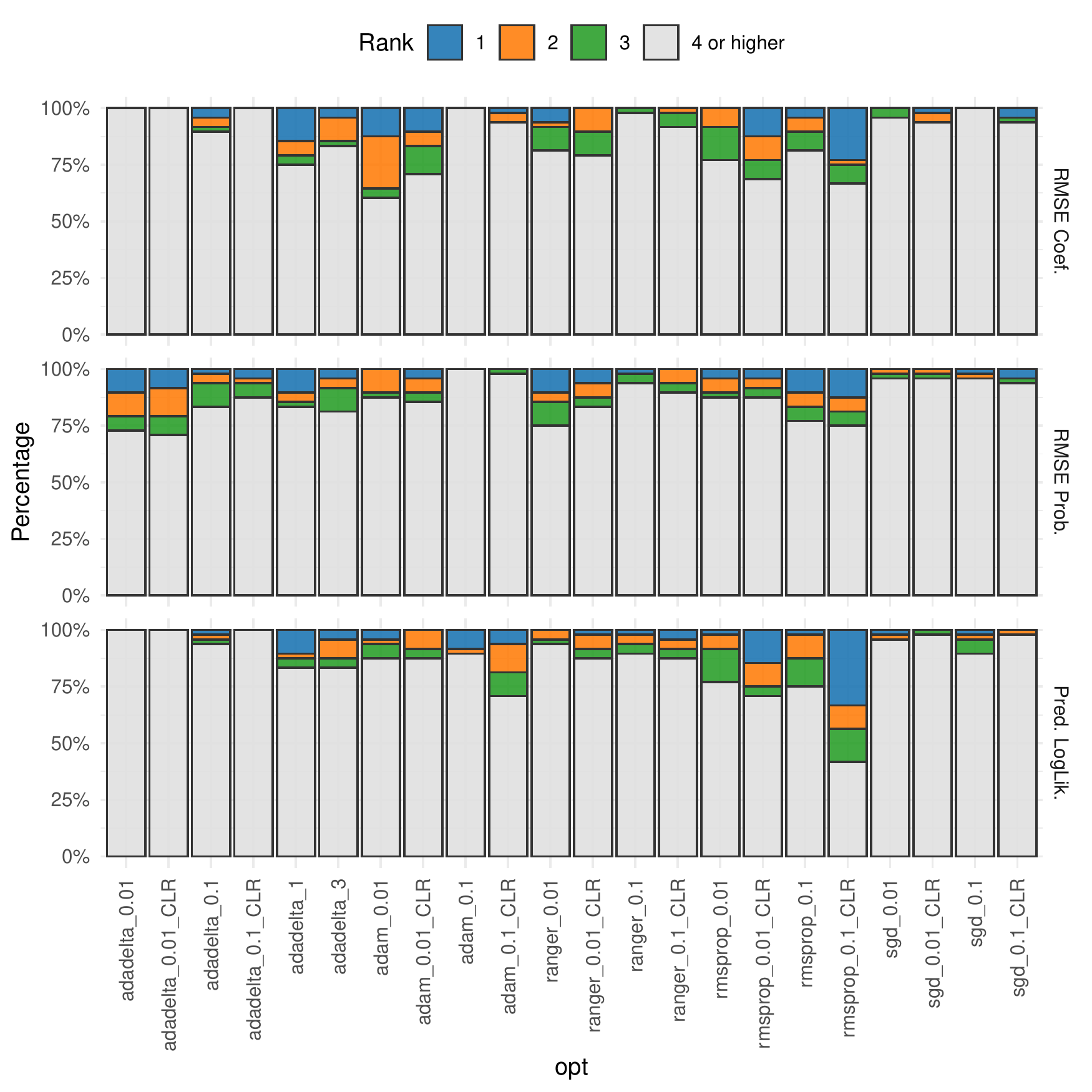}
    \caption{Comparison of various optimizers across simulation settings studied in experiment \ref{sec:comparisonEM}.
             Ranks were computed on performance averaged over 3 repetitions.}
    \label{fig:optimizers}
\end{figure}

NMDR introduces few hyperparameters that can be adjusted in order to improve results. The main hyperparameter introduced is the \textit{optimizer} used to optimize the neural network weights. Various optimizers that build on stochastic gradient decent (sgd) have been proposed and studied in recent literature such as \textit{rmsprop} \citep{rmsprop},  \textit{adadelta} \citep{adadelta}, \textit{adam}\citep{adam} and \textit{ranger} \citep{ranger}.
Those optimizers in turn often come with hyperparameters such as their learning rate which needs to be adjusted. In order to provide insights into the various optimizers' performance, we conduct a small benchmark study to assess the influence of the choice of an optimizer and to find a good default. Figure \ref{fig:optimizers} contains a comparison based on ranks for each optimizer across all 48 settings studied in experiment \ref{sec:comparisonEM}. \textit{rmsprop} with a learning rate of $0.1$ and a cyclical learning rate schedule leads to the best average rank ($6.55$).
As EM based methods are often restarted in order to prevent beeing stuck in poor local minima, we also study a setting where NMDR is restarted $3$ times, each time with a different optimizer.
We also determine a diverse combination of optimizers by adding \textit{adadelta} with a learning rate of 0.1 and \textit{ranger} with a learning rate of 0.01, both without CLR in order to determine whether restarts with different optimizers can improve over a single restart.
\textbf{Results}:
The \textit{rmsprop} optimizer achieves the lowest overall rank, yet no clear optimizer emerges as the best across all scenarios.
Therefore we suggest the rmsprop optimizer as a default setting when optimizing relatively simple settings such as the one proposed in experiment \ref{sec:comparisonEM}, and to tune the optimizer and its learning rate in case more computational resources are available.

\end{appendix}

\end{document}